\documentclass[pdflatex,sn-mathphys]{sn-jnl}

\usepackage{comment}
\usepackage{subcaption}%
\usepackage{colortbl}

\usepackage{balance}
\usepackage{soul}
\usepackage{booktabs}
\usepackage{graphicx}
\usepackage{amstext}
\usepackage{multirow}
\usepackage{multirow, makecell}
\usepackage{hyperref}

\usepackage[boldmath]{numprint}
\algnewcommand\algorithmicforeach{\textbf{for each}}
\algdef{S}[FOR]{ForEach}[1]{\algorithmicforeach\ #1\ \algorithmicdo}

\npdecimalsign{.}
\nprounddigits{3}

\jyear{2022}%

\theoremstyle{thmstyleone}%
\theoremstyle{thmstyletwo}%

\theoremstyle{thmstylethree}%

\begin{document}

\title[Multi-Value Alignment in NorMAS]{Multi-Value Alignment in Normative Multi-Agent System: An Evolutionary Optimisation Approach}


\author*[1]{\fnm{Maha} \sur{Riad}}\email{maha.riad@ucdconnect.ie}

\author[2]{\fnm{Vinicius} \sur{Renan de Carvalho}}\email{viniciusrenan.carvalho@gmail.com}

\author[1]{\fnm{Fatemeh} \sur{Golpayegani}}\email{fatemeh.golpayegani@ucd.ie}

\affil[1]{\orgdiv{School of Computer Science}, \orgname{University College Dublin (UCD)}, \orgaddress{\street{Belfield}, \city{Dublin}, \postcode{D04 V1W8}, \state{Dublin 4}, \country{Ireland}}}

\affil[2]{\orgdiv{Escola Polit\'ecnica (EP)}, \orgname{University of Sao Paulo (USP)}, \orgaddress{\street{Rua da Reitoria}, \city{S\~ao Paulo}, \postcode{05508-220}, \state{Butantã}, \country{Brazil}}}




\abstract{
Value-alignment in normative multi-agent systems is used to promote a certain value and to ensure the consistent behavior of agents in autonomous intelligent systems with human values.
However, the current literature is limited to incorporation of effective norms for single value alignment with no consideration of agents' heterogeneity and the requirement of simultaneous promotion and alignment of multiple values. 
This research proposes a multi-value promotion model that uses multi-objective evolutionary algorithms to produce the optimum parametric set of norms that is aligned with multiple simultaneous values of heterogeneous agents and the system.
To understand various aspects of this complex problem, several evolutionary algorithms were used to find a set of optimised norm parameters considering two toy tax scenarios with two and five values are considered. The results are analysed from different perspectives to show the impact of a selected evolutionary algorithm on the solution, and the importance of understanding the relation between values when prioritising them.  
}



\maketitle

\section{Introduction}\label{sec1}
Multi-agent systems (MAS) can model complex applications such as intelligent transport systems \cite{ghanadbashi2022using} and energy systems \cite{golpayegani2016multi}, using agents with certain objectives, who can take actions to achieve them autonomously. Such agents interact with the environment, and can compete or collaborate with one another. One of the main contributors that assist in the collaboration process is to coordinate the agents' behaviour by unifying their norms and using Normative Multi-agent systems (NorMAS).

The norms in NorMAS are regulative norms defined by a social group or the majority of agents to regulate their social behaviour~\cite{riad2022run}. For example, in a traffic system, in most countries the norm is to give priority to emergency vehicles. Another example, it is a norm that passengers leave front seats in buses to old people to be close to the doors. These examples represent guidelines that might be recommended in some societies, obligated, or just non forbidden (given permission) in others ~\cite{riad2022run}, so when the agents are aware of the norms of a specific environment they are operating in, they can synchronize their behaviour with other agents, facilitate group decision making and collaborate. Moreover, promoting human values in MAS is crucial to reflect real-applications.

In the context of this research, the term 'value' refers to motivational values, which represent standards that serve desirable objectives \cite{sierra2019value}. In other words, a 'value' will represent a preferred state or behaviour \cite{sierra2019value}, such as equality, health, wealth, happiness, fairness, etc. \cite{bench2017norms}. For example (to differentiate between a 'norm' and a 'value'), as a norm, companies give their employees maternity leaves if they have new born baby. However, if the values of one of the companies support equality between men and women, both can have equal maternity leaves \cite{sierra2019value}. 

The concept of value-alignment was introduced in \cite{originalProblem},\cite{serramia2020qualitative} and \cite{sierra2019value}, to reflect alignment of the behavioural norms of the intelligent agents with shared human values. Researchers used several techniques trying to address this challenge, these techniques included: reasoning strategies \cite{bench2017norms}, learning methodologies \cite{noothigattu2019teaching}, utility-based approaches\cite{heidari2019agents}, and genetic algorithms \cite{originalProblem}. However, the proposed solutions are only limited to one or more of the following points. First, matching the norms (or sub-set of norms) with only one value or with the most preferred sub-set of values, while in the real word all the values regardless of their internal compatibility need to be aligned with the whole set of norms imposed by a society. Second, these models do not consider heterogeneous multi-agent environments in which different groups of agents might support different values, especially when these values are incompatible. For example, supporting both fairness and equality may be conflicting, as ensuring fairness does not necessary support equality. Third, some works directly derive norms from values of the system. However, in many systems  norms and values may be incompatible and they should be considered independently. For instance, a community can have a value of supporting equality, at the same time have a norm of giving priority to senior people in queues, or exempting them from paying tax.

In this paper, we address these limitations by proposing a \textbf{N}ormative \textbf{A}lignment and \textbf{O}ptimisation Model (NOA), that has three main goals:
\begin{enumerate}
    \item Choose the \textit{best set of norms} in NorMAS with heterogeneous group of agents
    \item \textit{Optimize multiple values} in NorMAS, these values can be:
        \begin{itemize}
            \item compatible and \textit{incompatible} vales
            \item defined by \textit{heterogeneous} groups of agents
        \end{itemize}
    \item Align independent sets of norms and values in NorMAS
   
\end{enumerate}
To reach these goals,  we formalised the problem as a multi-objective optimization problem (MOP), in which we represented the values by objectives that need optimization, and modeled the norms as the decision variables. This allow us to get the \textit{best set of norms} (the decision variables) with respect to the values (the objectives), and so, align the norms and the values. Moreover, solving it as a multi-objective optimisation problem, (i) allows the system to facilitate \textit{Optimising multiple values} defined by \textit{heterogeneous} groups of agents, and (ii) allows multiple compatible and \textit{incompatible} values (objectives) \textit{optimisation}.

We proposed to solve this problem using multi-objective evolutionary algorithms (MOEAs) as they have been successfully applied to solve MOPs~\cite{surveyMetaHeuristics} in several domains including logistics, ride-sharing~\cite{deCarvalhoFatemeh2022},  environmental/economic dispatch (EED) problems~\cite{qu2018survey}, feature selection for machine learning problems \cite{coelho2021review}, and by optimising antibiotic treatments~\cite{ochoa2020multi}. 
We applied several MOEAs (NSGA-II, MOEA/DD, SPEA2, and MOMBI2) on different evaluation scenarios to analyse the performance of each of the MOEAs.

Accordingly, our proposed model, is a multi-value promotion model that uses multi-objective evolutionary algorithms to produce an optimum parametric set of norms that is aligned with shared system values as well as multiple simultaneous values of heterogeneous agents groups (NAO). NAO optimises a synthesised set of norms in a multi-value normative multi-agent system using different evolutionary algorithms for optimisation and is evaluated using different scenarios that measure the effect of using different combination of values. Our contribution is three-fold:
\begin{itemize}
    \item Multiple values alignment: we show the capability of choosing the optimum values for a parametric norms set while aligning it with a set of multiple optimised values.
    \item Incompatible and compatible values alignment: we model the problem as a multiple-objective problem to enable the simultaneous optimisation of all values regardless of their compatibility.
    \item Heterogeneous agents groups' values alignment: we align values from different heterogeneous groups of agents while considering shared system values.
\end{itemize}
The rest of this paper is organized as follows. Section \ref{sec:relatedWork} covers the related work addressing the value-alignment challenge, and the  background for MOEAs. Section \ref{section3}, covers the problem formulation, hypothesis and assumptions, and introduces the tax toy scenario that is used for the illustration of our model and in the evaluation. Section \ref{sec:Model}, the proposed model NAO is discussed and then evaluated in section \ref{sec:experiments}. Section \ref{sec:Discussion}, discusses how we reached the hypothesis and tackled the identified challenges in the literature. Finally, the paper is concluded in section \ref{sec:conclusion}.
\section{Related Work}\label{sec:relatedWork}
This section reviews the most recent work in value-alignment in NorMAS and their gaps. It then presents the background for Multi-Objectives Evolutionary Algorithms (MOEAs) as they will be the base of our proposed solution to overcome these gaps in section \ref{sec:Model}.
\subsection{Value-Alignment in Normative Multi-Agent Systems}

Contributions towards value-alignment can be classified mainly under the three following directions.

\textbf{Utility-based approaches:} the problem of selecting the best subset of norms from a given set of norms based on values was encoded as a linear program in \cite{serramia2018exploiting}. The authors used constraints to avoid conflicting norms, and modeled values as objectives to be optimised. The authors claim to construct the best set of norms relying on values as well as preferences of the society. Based on that work, the linear program was used in a decision support system for taking decision of approving propositions submitted to the French National Assembly by French citizens who wish to contribute to law construction \cite{serramia2018moral}. While \cite{serramia2018exploiting, serramia2018moral} relies on a quantitative approach that chooses norms that maximises a utility function, which can prioritize values aligned with larger number of norms over the most preferred values, \cite{serramia2020qualitative, serramia2021dominant} use a qualitative approach for choosing the right set of norms. This approach is based on formulating a ranking norm system according to the values preferences. 

Opposite to the previous approaches that select the right subset of norms from a given set of norms, by choosing norms that best support the most preferred values by the society, \cite{originalProblem} find the value that has the best performance with a predefined set of parametric norms. In the proposed model, the best parametric values of norms are chosen based on a given value, using a value-guided parametric norms synthesis methodology. For assessing the effect of individual norms on the alignment process,  Shapley Values concept is introduced. The compatibility between different values and the extent to which NorMAS can be aligned with them are assessed. Although this work has shown interesting results in value-alignment and promotion of values in NorMAS by formulating the problem as an optimisation problem, the important gap that still exists and well highlighted as their future work is promotion of multiple values simultaneously, while the compatibility of the norms and values are considered. It is important to note that in real world applications, we cannot only consider one shared value for all the entities in the system; as there may be multiple shared system level values as well as other group level (e.g., community level) values. This will introduce the heterogeneity of values that might not be necessarily compatible with each other and thus it will be challenging to propose a solution (i.e., optimised parameters for the norm set) that takes into account the values. 

In \cite{riad2022normative, riad2022run}, the set of norms is not primary defined at design time like the previous methods (i.e. the set of norms is initially empty), but it is synthesised(created and formulated) at run-time. The values are modelled as objectives to formulate a utility function that the system shall use in norms reasoning in case of conflicting norms. 

\textbf{Learning mechanisms:} the authors in \cite{rodriguez2022instilling} addressed the value-alignment problem using Multi-objective reinforcement learning; by adding the ethical knowledge to the reward function. They defined a value as a paired tuple consisting of: a finite set of norms promoting certain value, and an evaluation function that measures the impact of certain action on promoting that value. However, there proposed methodology works for a single agent promoting a single value. 

The value-alignment challenge in \cite{nahian2020learning} has a different focus than this paper, as the authors use the values directly to construct the norms. They use educational comic strip for children for training several machine learning algorithms to learn the main values from the main character of the stories, then classify situations as normative and non-normative. However, in real world, it should not be assumed that norms and values are primarily aligned. In this paper, we investigate how to align independent set of norms and values. Also, \cite{noothigattu2019teaching} defines the ethical values as norms, and makes an agent learn them from expert demonstrations using inverse reinforcement learning. In parallel, they use reinforcement learning for defining the policies that serve the environment rewards, and use contextual bandit-based orchestrator for choosing the best policy. This approach is limited to selecting only one policy at a time, so there is no coordination between the norms and the other system objectives simultaneously. 

The authors of \cite{heidari2019agents} created norms to promote certain values and directly linked them in a value tree, by statically defining the links between the norms and appropriate actions. They defined values for heterogeneous groups of agents with different norms, and each agent follows the norms applied by the majority of its social group. However, their technique of norms creation in the first place needs to be more generic as it is currently identified at design time.

\textbf{Reasoning techniques:} Nova \cite{aydougan2021nova}, a framework for agent-based automatic negotiation is proposed for norms revision. In Nova, agents’ values are used in the process of norm revision by the aid of ontology-based reasoning. When an agent initiates an offer for the other agent to negotiate, a scoring function is used by each to analyse the norms and then they create new norm that is aligned with both agents’ values, but it is not necessary to be the best norm. The evaluation of Nova involved the negotiation of human participants against the implemented agent. However, the use of this technique means that the agents should have a reasoning capability.
\begin{sidewaystable}
\sidewaystablefn%
\begin{center}
\begin{minipage}{\textheight}
\caption{Related Work}\label{table:RelatedWork}
\begin{tabular}{m{0.75cm}m{3.8cm}m{2.7cm}m{4cm}m{6cm}}
\toprule%
Paper&Technique&Case Study&Advantage&Limitation\\
\midrule
\cite{serramia2018exploiting,serramia2018moral}&Multi-objective-optimisation \& Utility-based approach&Border control at an international airport& Handle conflicting norms using constraints \& the set of norms is independent of the set of values &Limited to the problem formalization \& domain specific \& Do not consider heterogeneous agents' norm\\
\hline
\cite{serramia2021dominant,serramia2020qualitative}&Qualitative decision making using preferences&Border control at an international airport&Multiple values alignment with multiple norms& Apply compatible norms only \& do not consider heterogeneous agents norms\\
\hline
\cite{originalProblem}& Genetic Algorithm& Taxation Scenario & Align an independent set of Norms for heterogeneous agents and a value & Align a single value\\
\hline
\cite{riad2022run, riad2022normative}&Utility function \& case-based reasoning&Traffic junction&Align multiple values with multiple norms&Do not consider aligning heterogeneous agents values; all the values are the system's values\\
\hline
\cite{rodriguez2022instilling}&Multi-Objective reinforcement learning&A public civility problem&coordinating the agent's individual objectives& Used single agent \& single value\\
\hline
\cite{nahian2020learning}&Machine learning&Children’s educational comic strip &Classification of normative and non-normative behaviour based on values&They do not distinguish between the concept of norms and values and just treat norms as values\\
\hline
\cite{noothigattu2019teaching}&Inverse reinforcement Learning \& reinforcement learning&Pac-Man video game&Policy learning based on norms, values and environmental rewards&It selects a single policy only at a time using a contextual bandit orchestrator rather than having an aggregated policy \& applied on a single agent\\
\hline
\cite{heidari2019agents}&Utility-based decision making&Public donations&Agents have heterogeneous values&Norms are created for promoting certain values (i.e. they are not independent sets)\\
\hline
\cite{aydougan2021nova}&Ontology-based reasoning&Patient privacy and national security trade-off&agents’ values are used in the process of norm revision&Require reasoning capabilities\& the agents are not heterogeneous\\
\botrule
\end{tabular}
\end{minipage}
\end{center}
\end{sidewaystable}
\subsubsection{ State-of-the-art Discussion}

Most of the utility-based approaches enabled optimising multiple values (check Table \ref{table:literatureComp}) which is essential to support complex applications with multiple compatible and incompatible values. However, all of them except \cite{originalProblem} synthesise the same set of norms for all agents, while in real applications agents would need different norms  if they belong to heterogeneous groups. For example, the norms of wealthy group of people is different from the norms of poor people in aspects related to spending money. Although \cite{originalProblem} overcomes this by defining parameterized norms that enable each heterogeneous group to have its own norm, its model can align with only a single value at a time.

While \cite{noothigattu2019teaching} is able to learn norms and values simultaneously with the individual objectives, its main problem, as well as the other learning techniques \cite{rodriguez2022instilling,nahian2020learning,heidari2019agents}, is defining the norms directly based on the values, and not differentiating between the two concepts (norms and values) by using independent set for each of them. However, in real applications the norms and values are not necessarily linked, for instance a government may support equality value, but has the norm to exempt old citizens from paying taxes.

Also, despite that the negotiation mechanism introduced in \cite{aydougan2021nova} is able to align multiple values and separate the concept of norms and values, it does not ensure that the agreed on norm by the two negotiating agents is the optimum norm (i.e.belongs to the best norms sub-set). 

According to Table \ref{table:literatureComp} we can conclude that none of the illustrated related works is able to compromise multiple values while selecting the best set of norms (which are aligned with these values) in a heterogeneous NorMAS.
\begin{table}[!htbp]
	\caption{Related Work Comparative Analysis}
	\label{table:literatureComp}
	\centering
	\begin{tabular}{m{1cm}m{0.8cm}m{1.8cm}m{2cm}m{2cm}m{1.8cm}}
		\toprule
Approach&Paper&Multiple Values Alignment& Heterogeneous Norms Alignment & Independent Sets of Norms \& Values& Multi-Agent System\\
		\midrule
Utility-based&\cite{serramia2018exploiting,serramia2018moral}&\checkmark&X&\checkmark&\checkmark\\
&\cite{serramia2021dominant,serramia2020qualitative}&\checkmark&X&\checkmark&\checkmark\\
&\cite{originalProblem}&X&\checkmark&\checkmark&\checkmark\\
&\cite{riad2022run, riad2022normative}&\checkmark&X&\checkmark&\checkmark\\
Learning&\cite{rodriguez2022instilling}&\checkmark&X&\checkmark&X\\
&\cite{nahian2020learning}&\checkmark&X&X&\checkmark\\
&\cite{noothigattu2019teaching}&X&X&\checkmark&X\\
&\cite{heidari2019agents}&\checkmark&\checkmark&X&\checkmark\\
Reasoning&\cite{aydougan2021nova}&\checkmark&X&\checkmark&\checkmark\\
		\bottomrule
	\end{tabular}
\end{table}
\subsection{Multi-Objective Evolutionary Algorithms (MOEAs)}
MOEAs are heuristic techniques that provide a flexible representation of the solutions and do not impose continuity conditions on the functions to be optimised. Moreover, MOEAs are extensions of EAs for multi-objective problems that usually apply the concepts of pareto dominance~\cite{pdfRefOlacir}.
In pareto dominance, a certain solution $sl_a$ in the decision space of a MOP is superior to another solution $sl_b$ if and only if $f(sl_a)$ is at least as good as $f(sl_b)$ in terms of all the objectives and strictly better than $f(sl_b)$ in terms of at least one single objective. Solution $sl_a$ is also said to strictly dominate solution $sl_b$~\cite{pdfRefOlacir}.
Many of MOEA algorithms are genetic algorithms including \textbf{NSGA-II}~\cite{nsgaII}, \textbf{SPEA2}~\cite{spea2} and \textbf{MOMBI2}\cite{mombiII} that differ from each other mainly in the way that solutions are ranked at every iteration~\cite{vazquez2012mixture}, or in their decomposition technique (e.g., in  \textbf{MOEA/DD}) \cite{moeaddd}. 
\section{Problem Formulation and Toy Scenario} \label{section3}
Let us consider a heterogeneous normative multi-value multi-agent system that is composed of a finite set of regular agents as $Ag=\{ag_1,ag_2,...,ag_n\}$. Each agent $ag_i$ has a set of values $V_{ag_{i}}$, a set of properties $Pr_{ag_{i}}$, a set of actions $A_{ag_{i}}$, and a set of adopted norms $N_{ag_{i}}$. There is one regulative agent $r$ that is responsible to synthesise the norms set $N$, in which $N_{ag_{i}} \subseteq N$. The norms are parametric norms, i.e. each norm $n_j$ has a set of parameters $P_{n_{j}}$ that can contain unbounded or constrained elements with discrete or continuous domains. The regulative agent $r$ has a set of values $V_{r}$ as well.
In each step (iteration), each regular agent $ag_i$ performs actions from $A_{ag_{i}}$ and applies its set of adopted norms $N_{ag_{i}}$. The regulative agent also applies actions chosen from its set of actions $A_{r}$. Corresponding to the agents' new situations, a global state $s$ is captured by $r$.
In such a system, $r$'s  main challenges are: to synthesise the optimum set of norms that ensures the alignment of its own values $V_{r}$ and each of the regular agent's values $V_{ag_{i}}$ (which is shared between a subset of agents), and to optimise the synthesised set of norms even in case of incompatible values. 
\subsection{Hypothesis} Formulating the multi-value alignment problem in NorMAS as a multi-objective optimization problem and solving it using Multi-Objective Evolutionary Algorithms will allow \textbf{optimizing multiple values (compatible and incompatible) while selecting the best set of norms that are aligned with these values in heterogeneous NorMAS, in which different groups of agents have different values}.
\subsection{Assumptions and Limitations}
The model is constructed based on the following assumptions:

\begin{itemize}
    \item $g_{k_{ag_{i}}}$ defines the agent $ag_{i}$ group, and $Gr$ defines a finite set of heterogeneous groups.
    \item Initially, the number of agents in each $g_k \in Gr$ is equal.
    \item The agents of the same group $g_k$ have the same set of values $V_{g_k}$.
    \item There is a regulative agent $r$ that is aware of the set of values $V_{g_k}$ of all of the groups.
    \item Agent $r$ gives equal priority to all the sets of values in the system.
    \item All values are formalised in the form of minimisation or maximisation objectives.
    \item All the sets of values are predefined, finite and static.
\end{itemize}
 \subsection{Tax System Scenario}\label{taxScenario}
\begin{figure*}[!htbp]
\centering
\centerline{\includegraphics[scale=0.3]{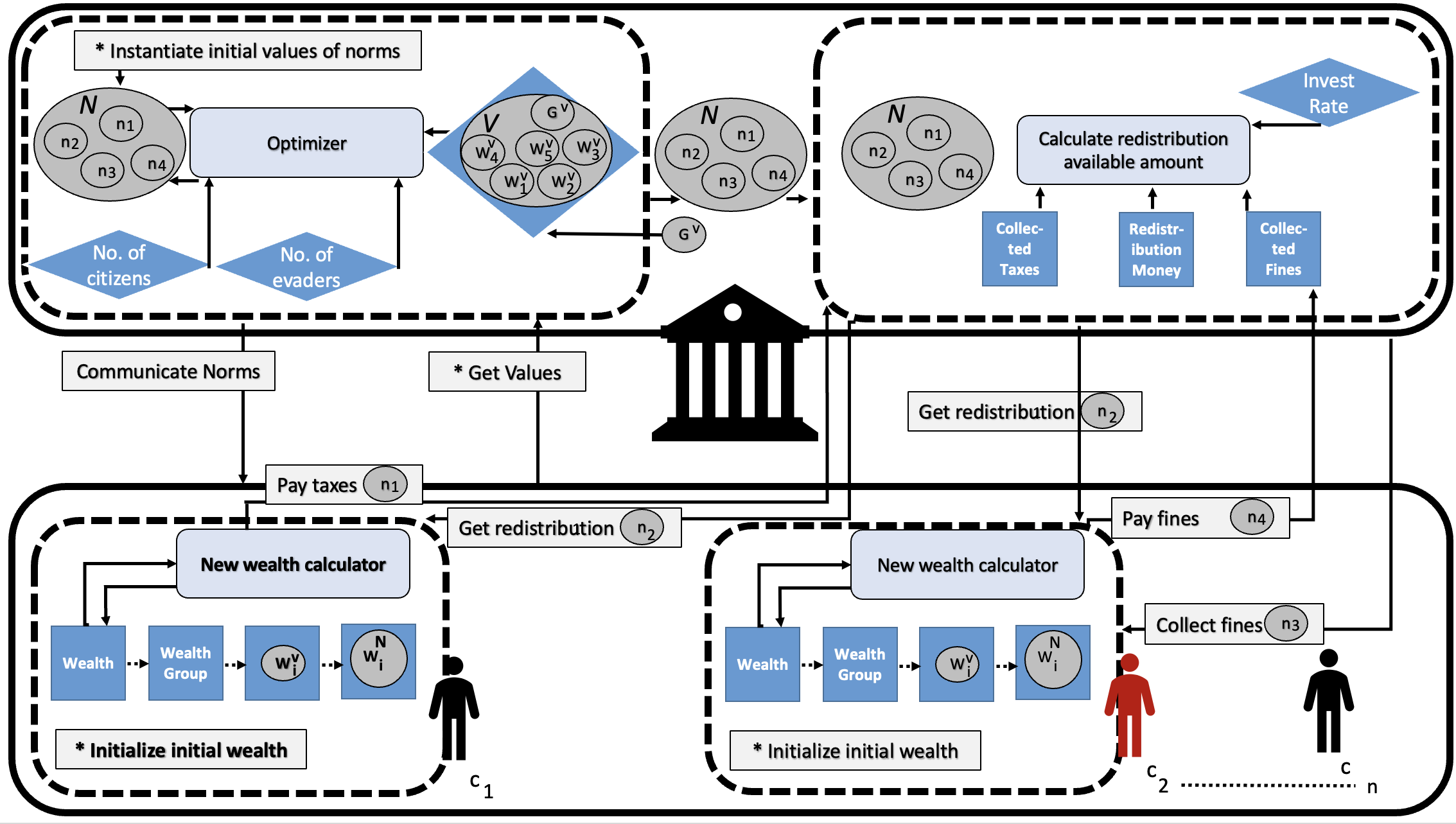}}
\caption{Norm Synthesis Optimisation in a Multi-Value Normative Multi-Agent System Conceptual Model}
\label{fig:ConceptualModel}
\end{figure*}

 As a running example and for evaluation we will use an adapted tax system toy scenario introduced in \cite{originalProblem}. In this scenario, the regular agents set $Ag$ will represent the set of citizens and the regulating agent $r$ will represent the government. The main idea of the system is that the government collects taxes from the citizens according to their wealth group. $collects_taxes$ is an action which belongs to $A_{r}$, actions are represented in Fig. ~\ref{fig:ConceptualModel} by rectangles.There are five wealth groups, the $1^{st}$ group represents the poorest group while the $5^{th}$ group represents the richest group. A percentage of the citizens do not pay taxes and will be considered as evaders. However, if they were caught by the government they will be punished and will pay the evaded payment in addition to extra fines. In case they do not have sufficient funds only the available money is collected to avoid getting the citizen into debt. After the taxes and fines are collected a 5\% will be considered as a fixed interest rate that is added to the total collected amount. Then, the total collected money $cr$ will be redistributed back to the citizens depending on their wealth group. Initially (i.e. before simulation), The wealth of each citizen is randomly assigned after being initialized using a random uniform distribution $U$(0,100). Then, agents are allocated to their corresponding wealth group, with a constraint that the wealth groups have equal number of citizens. The main characteristics of the system are as follows. First, each of the citizens  has four main properties in their properties set, which describes its current state. The properties are:

 \begin{itemize}
     \item Wealth $(w_i)$: it has a numerical value that represents the amount of money citizen $i$ currently has.
     \item Wealth group $(g_k)$: it represents the wealth group the citizen belongs to according to its wealth $w_i$. 
     \item Evader flag $(e)$: it reflects whether this citizen is evader and will not pay taxes or not. 
     \item Primary Wealth $(pw_i)$: it has a numerical value that represents the wealth of the citizen $i$ at the begging of a time-step before taking any action and before its state change.
 \end{itemize}
 
 Second, each citizen has a set of values $V_{ag_{i}}$, for simplicity in this example, citizens in the same wealth group have the same fixed set of values. In other words, each wealth group has a set of values $V_{g_{i}}$, this could represent the community values. Only it is assumed that wealth group $g_2$ does not have a value to simulate citizens with no particular values in real word scenarios. Third, the government has its own set of values as well $V_r$ and a set of parametric norms, which has initial randomly defined values. The norms are defined in the same manner they are stated in \cite{originalProblem}:
  \begin{itemize}
     \item \textbf{n1} defines the tax rate $collect_j$ each wealth group is expected to pay at each time-step. The parametric set of the norm is defined as $P_{n_1}$ = $\{collect_j\}_{j=1,...5}$. The tax rate values are restricted between 0 and 1.
     \item \textbf{n2} defines the fractional percentage $redistribute_j$ each wealth group will take back from the redistribution amount at the end of each time-step. The parametric set of the norm is defined as $P_{n_2}$ = $\{redistribute_j\}_{j=1,...5}$, the values are between 0 and 1 and the sum of the fractions is constrained to be equal to 1.
     \item  \textbf{n3} defines the catch rate of evaders. This single parameter is defined as $P_{n_3}$ = $\{catch\}$. Its value is constrained to be between 0 and $1/2$ to reflect the difficulty of law-enforcement tasks.
     \item \textbf{n4} defines the extra fine defined as punishment when an evader is caught. This single parameter is defined as $P_{n_4}$ = $\{fine\}$. However, the total amount to be paid by a caught evader, which is equal to the fine plus the taxes amount, can not exceed the total wealth of the evader. 
 \end{itemize}
 
 The main challenge of this system is represented in the government responsibility to optimise the parameters' sets $P_{n_i}$ of the previously defined four norms belonging to $N=\{n_1,n_2,n_3,n_4\}$, while aligning them with the values of the government, as well as the regular citizens values. The values are defined as follows.
\begin{itemize}
    \item \textbf{Value 1 (Obj1)}: the value of the government is \emph{Equality}, which is calculated using equation \ref{eq:equality} introduced in \cite{originalProblem}. $GI(s)$ represents the Gini Index of the global state $s$. The Gini Index \cite{giugni1912variabilita} is an indicator of inequality, where $w_k$ is the wealth of agent $ag_k$ and $\overline{w}$ is the average wealth of all agents at state $s$.
    \begin{equation}\label{eq:equality}
        Equality=1-2.GI(s),\: with\: GI(s)=\frac{\Sigma_{i,j\in Ag}\mid w_i-w_j \mid}{2.\mid Ag \mid ^2.\overline{w}}
    \end{equation}

    \item \textbf{Value 2 (Obj2)}: the value of citizens in wealth group $g_3$ is \emph{Fairness}. The main aim of this value is to have the highest number of evaders in wealth group $g_1$ (the poorest group). To calculate the estimated chance of evaders in $g_1$ equation \ref{eq:fairness} is used as suggested in \cite{originalProblem}.
    \begin{equation}\label{eq:fairness}
        Fairness=2.C[g_i(s)=1\mid evader_i]-1
    \end{equation}
    \item \textbf{Value 3 (Obj3)}: the value of citizens in wealth group $g_5$ is to maximise their \emph{Wealth}. The main aim of this value is to have the maximum wealth portion from the total wealth. It represents the new wealth of the citizens after an iteration takes place. Equation \ref{eq:wealth} is used for calculating the new wealth.
    \begin{equation}\label{eq:wealth}
        Wealth=\frac{\Sigma_{i\in g_5}w_i}{\Sigma_{j\in Ag}w_j}
    \end{equation}
     \item \textbf{Value 4 (Obj4)}: the value of citizens in wealth group $g_4$ is to maximise the \emph{Gained Amount}. The aim of this value is to have the maximum gain portion from the common amount available for redistribution $cr$. The gained value is the difference between the citizen's new wealth $w_k$ and the old wealth $pw_k$ (Check the numerator in equation \ref{eq:gained}).
    \begin{equation}\label{eq:gained}
        Gained \: Amount=\frac{\Sigma_{i\in g_4}w_i-pw_i}{cr}
    \end{equation}
    \item \textbf{Value 5 (Obj5)}: the value of citizens in wealth group $g_1$ is related to the \emph{Collect Portion}. The aim of this value is to have the minimum  portion from the collect rate out of 1 (total portion of collect rates). To inverse this to a maximisation objective we have:  \ref{eq:collect}.
    \begin{equation}\label{eq:collect}
        Collect \: Portion=1-Collect_1
    \end{equation}
\end{itemize}

The best alignment between the synthesised set of parametric norms and the values is achieved by maximising these 5 values (objectives).

\section{Norm Alignment and Optimisation Model}\label{sec:Model}
 The main aim is to find the best values of the parametric norms when the agents' and system's values are satisfied (specifically optimised). Although, as defined in \cite{originalProblem}, the primary objective is to systematically calculate the parametric norms that maximally align with defined set of values by computing equation ~\ref{eq:normAlign}, they only align a single system value $v$ at a time.
\begin{equation}\label{eq:normAlign}
    N^*=arg_{N'\subseteq N}max \: Algn^{Ag}_{N', V}
\end{equation}
They search for the best values of the parametric norms while optimising $v$. They solve the problem using Genetic Algorithm (GA) \cite{luke2016population} and claim to reach the optimum value when exceeding a specific threshold or when finishing a given number of iterations. However, in order to aligning multiple values in this setting the problem should be considered as a multi-objective optimisation problem (MOP).

Multi-objective optimisation (MOPs) consists of finding solutions which simultaneously consider two or more conflicting objectives to be minimised or maximised \cite{nagymulti}. Thus, the search aims to find a set of solutions, each one reflecting a trade-off between the objectives. MOPs are formulated using: objective functions, constraints, decision variables and their bounds \cite{nagymulti}. 

Respectively, in NAO, we formulate the problem identified in Section \ref{section3} as a multi-objective optimisation problem. We define the agents' and system's values as the objective functions to be optimised, and the norms as the decision variables.

Afterwards, NAO uses different Multi-Objective Evolutionary Algorithms (MOEAs) to solve the multi-objective optimization problem.
\begin{figure*}[!htbp]
\centering
\centerline{\includegraphics[scale=0.4]{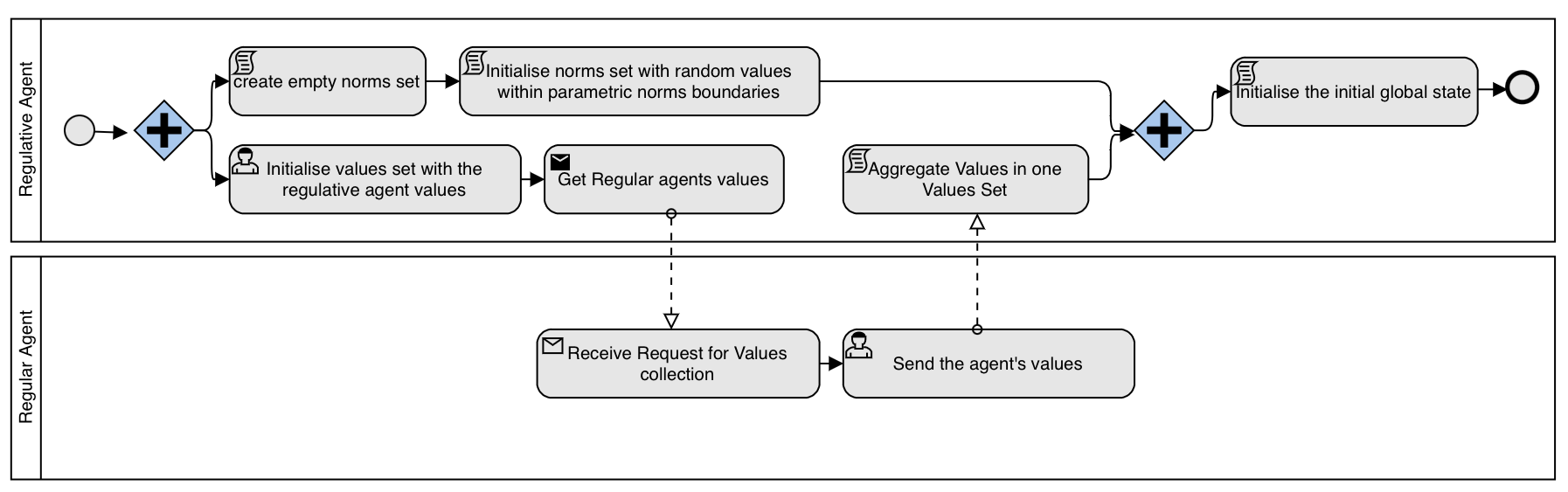}}
\caption{System Initialisation Process Instantiated by the Regulative Entity}
\label{fig:InitialisationProcessss}
\end{figure*}
Accordingly,NAO synthesises the best set of norms (decision variables) based on the agents' values (objective functions) optimisation, using the following off-line approach to synthesise norms based on the optimised values:

\textbf{NAO Initialisation parameters: }Initially, the parametric norms set $N$ is initialised with random values between the norms' specified boundaries by the regulative agent $r$ (as seen in Fig. ~\ref{fig:InitialisationProcessss}). Also, the primary values of the Properties $Pr_{ag_{i}}$ for each of the agents are defined. The regulative agent $r$ sends the initial set of norms $N$ to the agents. The values of the set of agents $Ag$ in the system and the regulative agent $r$ are consolidated by $r$ in one set of values $V$. These initial values are used to calculate the global state $s_0$. Then, the processed  $N$, $V_r$, $s_0$, and $MOEA$ (which is the multi-objective evolutionary algorithm that will be used for optimization) are used as input parameters to start the NAO strategy used in Algorithm 1. 

\textbf{NAO Strategy: }NAO takes the following main steps in each iteration $t$, see Algorithm 1:\begin{enumerate}
   \item Each of the agents in $Ag$ and the regulative agent $r$ carry out their actions while applying the relative norms to these actions. These actions produce new global state $s_t$. [Lines 3-6]
   \item The regulative agent $r$ performs its actions $A_r$ on $s_t$ considering the current norms $N$. [Line 7]
    \item The regulative agent $r$ uses the new global state to perform the optimisation process using a multi-objective optimiser $MOEA$ and produces the new set of norms $N$ based on the optimised set of values $V$. [Line 8]
    \item The the new set of norms $N$ is communicated to all the agents in $Ag$. [Line 9]
\end{enumerate}
\begin{algorithm}
\caption{NAO Strategy}
\begin{algorithmic} [1]\label{alog1}
  \State\textbf{Input:} $N$, $V$, $s_0$, $MOEA$
  \ForEach{t} 
       \ForEach{$ag_i \in Ag$}
        \State $s_t \Leftarrow ag_i.act(N_{ag_{i}}, A_{ag_{i}},s_t)$
       \EndFor
       \State $s_t \Leftarrow r.act(N,A_{r},s_t)$
       \State $N \Leftarrow r.optimize(s_t, V, N, MOEA)$
       \State $r.inform(Ag, N)$
   \EndFor
   \State $N^{*} \Leftarrow N $
\end{algorithmic}
\end{algorithm}
\subsection{NAO in the TAX System Scenario}
In the tax system scenario illustrated in section \ref{taxScenario}, NAO's goal is to find the values of the parametric norms n1, n2, n3 and n4 while optimising the values in $V$: equality, fairness, wealth, gained amount and collect portion. As it is seen in NAO's conceptual model in Fig. \ref{fig:ConceptualModel}, the system is divided to two main divisions, the divisions of the government and citizens. Evaders are represented in red as they have different set of norms and actions than the normal citizens. In this model, first, NAO randomly initializes the norms and the wealth of the citizens, and consequently they are assigned to their corresponding wealth groups. Second, the norms set $N$ is communicated by the government (the regulative agent) to the citizens. Third, the citizens start applying the different actions and their corresponding norms. So, the normal citizens will start paying taxes according to the rate of their wealth group defined by $n1$. Then the government will start catching the evaders according to the catch value defined in $n3$. The caught evaders will pay their taxes plus the fines determined using $n4$. Afterwards, the government calculates the total amount of money available for redistribution. Subsequently, each citizen receives their portion from redistribution according to the redistribution rate defined by $n2$. Then, the citizens calculate their new wealth and move to their new wealth groups. Forth, based on $s_t$, the government uses the optimiser to decide the new values of norms by optimising the five values in $V$. This cycle is repeated until NAO reaches a stopping condition that represents a satisfying level of the optimisation of the values in $V$.
\section{Experimental Evaluation}
\label{sec:experiments}

We evaluated four algorithms NSGA-II, MOEA/DD, SPEA2, and MOMBI2 on solving both two and five objectives problems. Both problems are based on the tax scenario defined in section \ref{taxScenario}. The two objectives problem includes value 1 (Equality) and value 2 (Fairness), and the five objectives problem includes all the values.  We used $200$ agents to represent the citizens, and a randomly chosen number of evaders in each iteration. The number of segments that represents the wealth groups was set as $5$. The invest rate was $0.05$. We used Monte Carlo Sampling during $5000$ iterations similar to \cite{originalProblem}, but in our case, Monte Carlo runs after a meta-heuristic complete execution. For this sampling the \emph{path} was defined as $10$. All meta-heuristic run for $500$ generations, and with the maximum population size of $100$ for two and $210$ for five objectives. For MOEA/DD we followed \cite{moeaddd} and set $Nr=1$, $\delta=10$ and probability as $0.9$. Regarding to evolutionary operators we followed~\cite{WFGPisa}, where the SBX Crossover and Polynomial Mutation were employed and setup with distribution set as $\{n_c=20\}$ and $\{n_m=20\}$ respectively with probabilities $\{p_c=0.9\}$ and $\{p_m=1/n_p\}$, where $n_p$ is the number of decision variables in the problem. 

Regarding implementation, NAO was coded using Java JDK 14 using jMetal 5.7~\cite{jMetal} and jMetalHyperHeuristicHelper \footnote{\href{https://github.com/vinixnan/jMetalHyperHeuristicHelper}{https://github.com/vinixnan/jMetalHyperHeuristicHelper}} for meta-heuristics and components.

We present and discuss our results from three different perspectives. First, Hypervolume and IGD+ averages are compared to understand the performance of different algorithms in this context. Secondly, we present how is the pareto Front for each of the meta-heuristics. Finally, we analyse the produced best solutions from the problem perspective.
\subsection{Hypervolume and IGD+ comparisons}
We employed Hypervolume \cite{jaszkiewicz2018improved}, and IGD+~\cite{IGDPlus} averages obtained from the 30 executions as the algorithms performance comparison criterion. This is necessary due to the fact MOEAs produce a set of non-dominated solutions, and direct comparisons between pareto Fronts are not considered the best approach to determine which algorithm performs better. For this purpose, first, for each problem (two and five objectives), we joined all results obtained by all algorithms, found the nadir point (worst found), necessary for Hypervolume calculation, and took the Non-dominated set in order to generate the \emph{known pareto Front ($PF_{known}$)}, necessary for IGD+ calculation. Then, we calculated Hypervolume and IGD+ for each of the executions and generated averages for both quality indicators for each algorithm. Finally, we compared these averages using Kruskal-Wallis as the statistical test with a confidence level of $99\%$. In order to perform this, we first identified which algorithm has the best average according to the quality indicator, thus, all the other algorithms are compared to the best, generating a set of \emph{p-values}. We define an algorithm tied statistically with the best when a given \emph{p-value} is superior to the significance level of $0.01$.

Table \ref{tab:2obj} presents meta-heuristic comparison for the two objective problems. Here the mean for 30 executions, standard deviation (\emph{std}), and  \emph{max} value among the executions are presented.

\begin{table}[!htbp]
	\caption{Hypervolume and IGD+ averages for two objectives, highlighted values means best results, bold values means statistically tied results with the best. For Hypervolume, higher values are considered the best, while for IGD+ smaller are preferred.}
	\label{tab:2obj}
	\centering
	\begin{tabular}{llrr}
		\toprule
		& Metric &  Hypervolume &      IGD+ \\
		\midrule
		MOMBI2 & mean &     0.030525 &  1.242683 \\
		& std &     0.090247 &  0.354033 \\
		& max &     0.425940 &  1.973713 \\
		MOEA/DD & mean &     \textbf{0.859607 }&  \textbf{0.103377} \\
		& std &     0.090352 &  0.068957 \\
		& max &     0.974304 &  0.263218 \\
		NSGA-II & mean &     \cellcolor[HTML]{DEDBDB}\textbf{0.904303} &  \cellcolor[HTML]{DEDBDB}\textbf{0.056089} \\
		& std &     0.084679 &  0.053044 \\
		& max &     0.999828 &  0.239230 \\
		SPEA2 & mean &     \textbf{0.862924} &  \textbf{0.096932} \\
		& std &     0.180627 &  0.143008 \\
		& max &     0.999379 &  0.704847 \\
		\bottomrule
	\end{tabular}
\end{table}

Regarding \emph{max} for Hyperolume, SPEA2 found the highest value. However, it also has the highest \emph{std}, which means this high value rarely occurs. NSGA-II has the best average, but when we consider both mean and std, we can see why MOEA/DD and SPEA2 results are statistically tied with NSGA-II. Regarding IGD+, the same situation happened, with NSGA-II as the best algorithm, but this time being the best one considering \emph{std}, \emph{mean} and \emph{max}. Finally, we can clearly see that all of these algorithms except MOMBI2 can be a good option for solving this two-objective problem.

\begin{figure}[!htbp]
	\centerline{\includegraphics[scale=0.25]{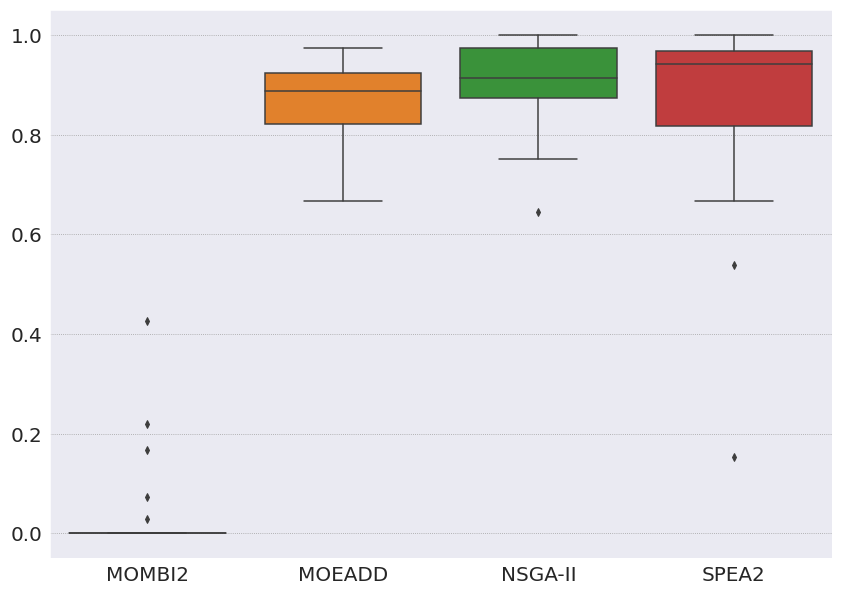}}
	\caption{Hypervolume box-plot for two objectives}
	\label{img:hypervolume2obj}
\end{figure}

\begin{figure}[!htbp]
	\centerline{\includegraphics[scale=0.25]{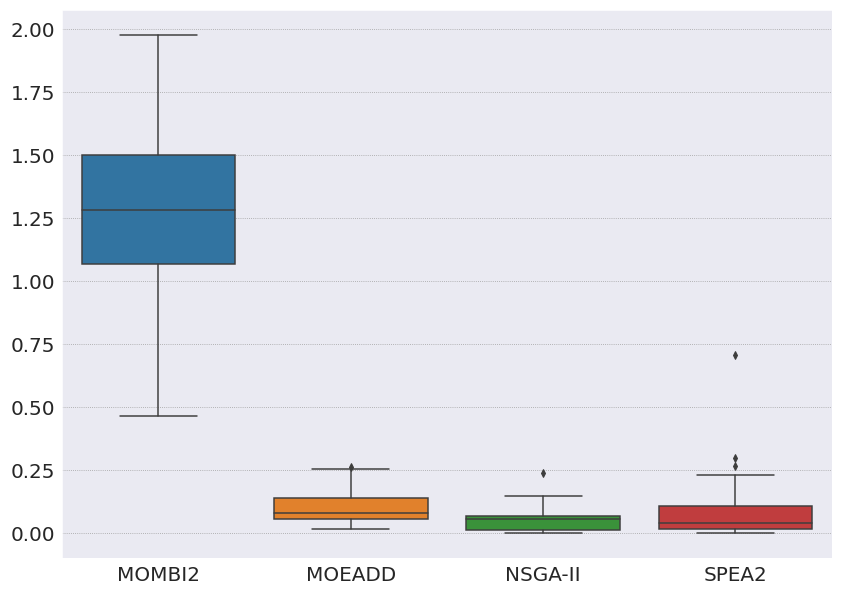}}
	\caption{IGD+ box-plot for two objectives}
	\label{img:igdp2obj}
\end{figure}

Figures \ref{img:hypervolume2obj} and \ref{img:igdp2obj} present, respectively, box-plots for Hypervolume and IGD+. Basically, these figures represent visually the same results shown in Table \ref{tab:2obj}. Here, we can see how MOMBI2 is the algorithm with more variance ( as shown in Table \ref{tab:2obj} regarding \emph{std}) while NSGA-II is the one with less variance. The reason for that is while NSGA-II usually found non-dominated solutions, MOMBI2 found dominated solutions making it sometimes having near to zero Hypervolume values. This is also clear when we analyse considering IGD+ where MOMBI2 always has bigger values.

Table \ref{tab:5obj} presents meta-heuristic comparison for the five objective problems. Here, the scenario is completely different. MOMBI2 was a terrigle algorithm for two objectives (see Table  \ref{tab:2obj}), but here it found the best \emph{max} value regarding to Hypervolume. In terms of standard deviation (\emph{std}) and \emph{mean} MOEA/DD is the best algorithm.  These results made MOMBI2 and MOEA/DD the best algorithms considering Hypervolume. However, in terms of IGD+, MOEA/DD stands as the best algorithms with a better IGD+ average with a statistical difference. Moreover, it also had the smallest \emph{std} and \emph{max} values.

\begin{table}[!htbp]
	\caption{Hypervolume and IGD+ averages for five objectives, highlighted values means best results, bold values means statistically tied results with the best. For Hypervolume, higher values are considered the best, while for IGD+ smaller are preferred.}
	\label{tab:5obj}
	\centering
	\begin{tabular}{llrr}
		\toprule
		& Metric &  Hypervolume &      IGD+ \\
		\midrule
		MOMBI2 & mean &     \textbf{0.378164} &  0.086026 \\
		& std &     0.022794 &  0.026758 \\
		& max &     0.412411 &  0.137319 \\
		MOEA/DD & mean &      \cellcolor[HTML]{DEDBDB} \textbf{0.386071} &   \cellcolor[HTML]{DEDBDB} \textbf{0.039012} \\
		& std &     0.009275 &  0.004931 \\
		& max &     0.402583 &  0.047243 \\
		NSGA-II & mean &     0.321709 &  0.109361 \\
		& std &     0.014935 &  0.014614 \\
		& max &     0.353739 &  0.145614 \\
		SPEA2 & mean &     0.343156 &  0.075612 \\
		& std &     0.009625 &  0.006417 \\
		& max &     0.361828 &  0.091765 \\
		\bottomrule
	\end{tabular}
\end{table}

\begin{figure}[!htbp]
	\centerline{\includegraphics[scale=0.25]{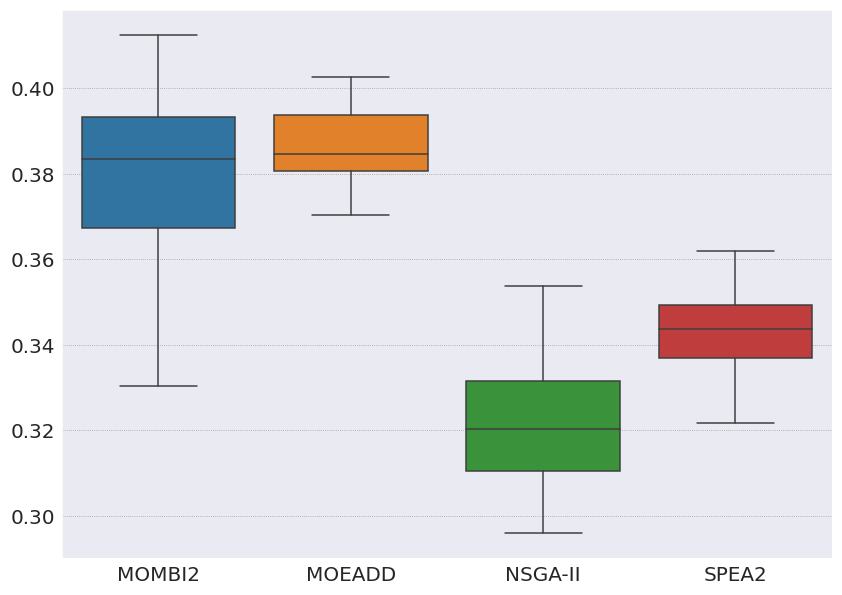}}
	\caption{Hypervolume box-plot for five objectives}
	\label{img:hypervolume5obj}
\end{figure}

\begin{figure}[!htbp]
	\centerline{\includegraphics[scale=0.25]{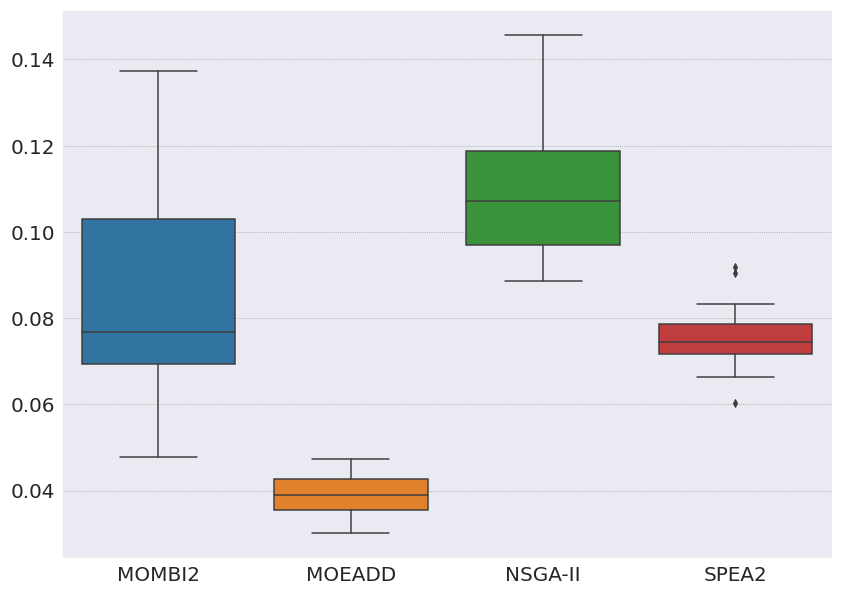}}
	\caption{IGD+ box-plot for five objectives}
	\label{img:igdp5obj}
\end{figure}

Figures \ref{img:hypervolume5obj} and \ref{img:igdp5obj} presents respectively box-plots for Hypervolume and IGD+ considering the five-objectives problem. Basically, these figures represent visually the same results shown in Table \ref{tab:5obj}. Here we can see, for Hypervolume, how MOMBI2 and MOEA/DD perform better than NSGA-II and SPEA2. However, unlike MOEA/DD, MOMBI2 has a big variance having the biggest Hypervolume values (considering one single execution) and minimal values smaller than NSGA-II and SPEA2 maximum values. MOEA/DD is more stable in terms of results, even not having the maximum value among the algorithms, which is the best choice for this problem. This is even more clear when we take into consideration IGD+ values where MOEA/DD is clearly the best performing algorithm in all aspects.

\subsection{pareto Fronts Analysis}
\label{subsec:paretoanalysis}

\begin{figure}[!htbp]
	\centerline{\includegraphics[scale=0.35]{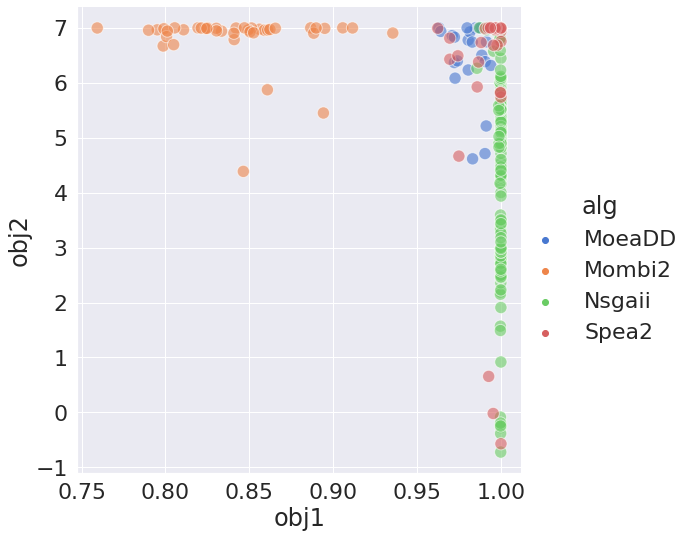}}
	\caption{pareto Front for algorithms at one execution considering two objectives}
	\label{img:2objpf}
\end{figure}

In this section, pareto Fronts generated by each algorithm are compared. Interactive graphs can be found in \href{https://htmlpreview.github.io/?https://github.com/vinixnan/PublicData/blob/master/NAO/Fronts/index.html}{\underline{\textbf{here}}}. Dominated solutions were also considered in order to provide a good view of how an algorithm can outperform others in terms of pareto dominance. Fig. \ref{img:2objpf} presents this for the two objective problem. Here we can see why MOMBI2 had bad Hypervolume and IGD+ results in the previous section due to the fact most of the solutions are dominated. NSGA-II also has a good amount of solutions. However, several are non-dominated, which means this algorithm has good results according to quality indicators.

\begin{figure}[!htbp]
	\centerline{\includegraphics[scale=0.14]{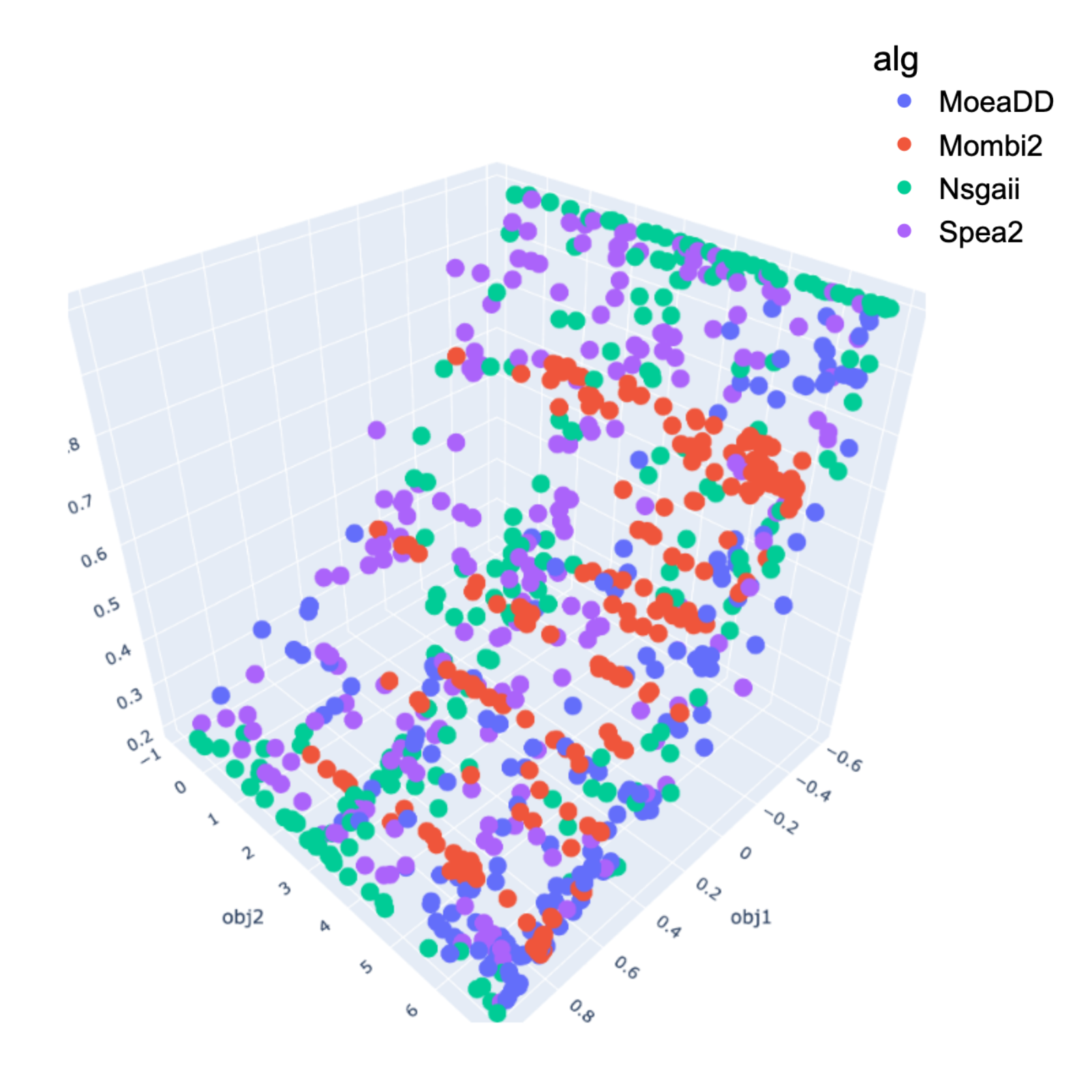}}
	
	\caption{pareto Front for algorithms at one execution considering objectives 1, 2 and 3}
	\label{img:iobj1_obj2_obj3}
\end{figure}

For the five-objective problem, we split the objectives into four groups by combining objectives one and two, which are used in the two objective problem, with one of the other objectives. For example, in Fig. \ref{img:iobj1_obj2_obj3}, objectives 1, 2 and 3 are shown. In this Figure, we can have a good view of the pareto shape, which is somehow linear and disconnected.  NSGA-II has solutions on extremes while MOEAD/D is more spread.

\begin{figure}[!htbp]
	\centering
	\begin{subfigure}[t]{0.5\textwidth}
	
			\centerline{\includegraphics[scale=0.15]{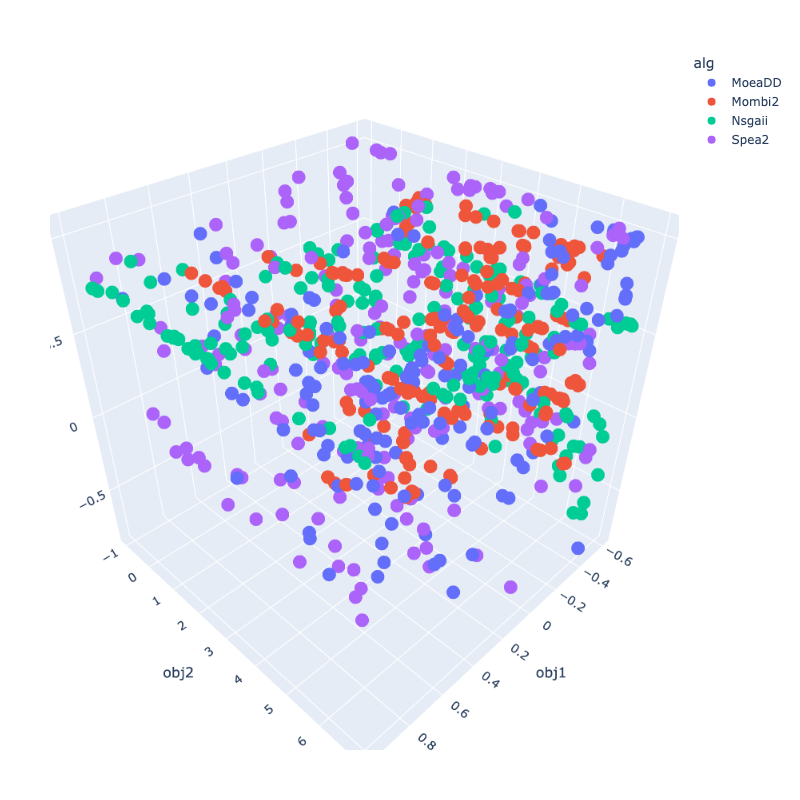}}

		\caption{1, 2 and 4 objectives}
		\label{img:iobj1_obj2_obj4}
	\end{subfigure}%
	~ 
	\begin{subfigure}[t]{0.5\textwidth}
		
			\centerline{\includegraphics[scale=0.15]{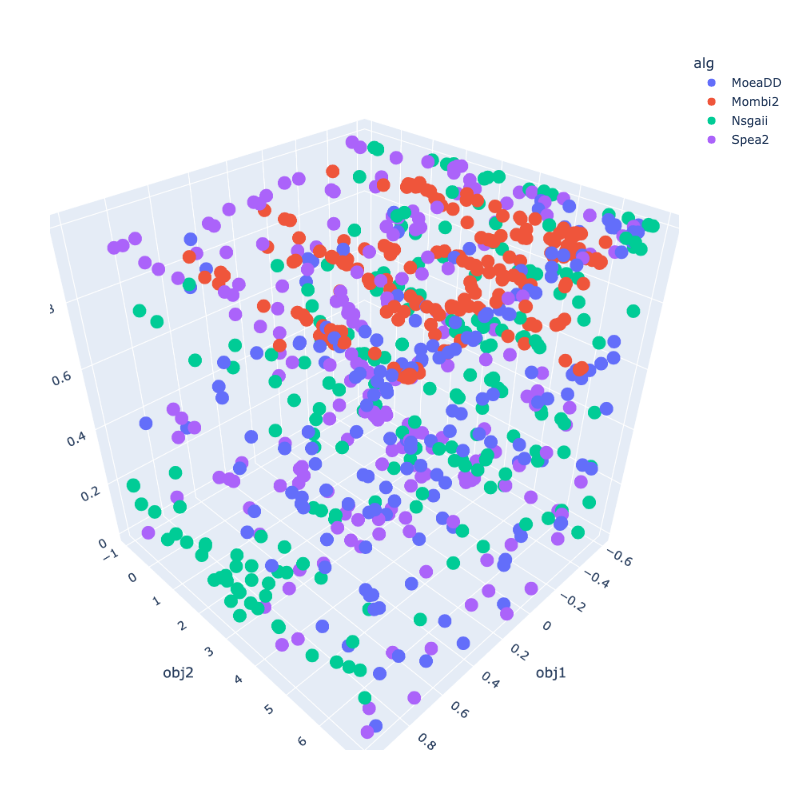}}
		
		\caption{1, 2 and 5 objectives}
		\label{img:iobj1_obj2_obj5}
	\end{subfigure}

	\begin{subfigure}[t]{0.5\textwidth}
		
			\centerline{\includegraphics[scale=0.15]{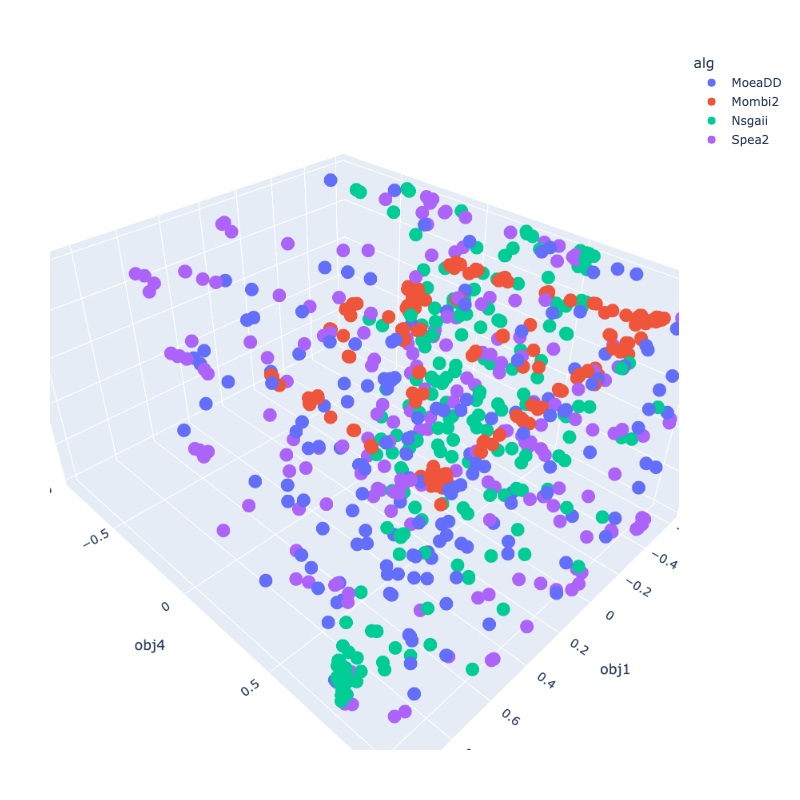}}
	
		\caption{1, 4 and 5 objectives}
		\label{img:iobj1_obj4_obj5}
	\end{subfigure}%
    ~ 
	\begin{subfigure}[t]{0.5\textwidth}
		
			\centerline{\includegraphics[scale=0.15]{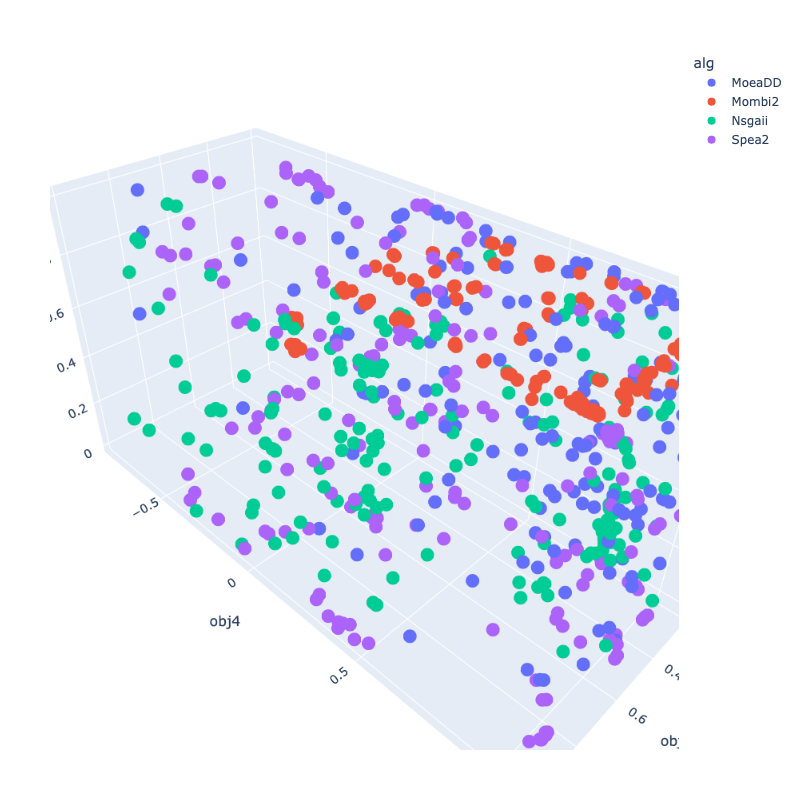}}
	
		\caption{3, 4 and 5 objectives}
		\label{img:iobj3_obj4_obj5}
	\end{subfigure}
	
	\caption{pareto Front for algorithms at one execution}
\end{figure}

Figures \ref{img:iobj1_obj2_obj4}, \ref{img:iobj1_obj2_obj5}, \ref{img:iobj1_obj4_obj5} and \ref{img:iobj3_obj4_obj5} presents 3D plots for other objectives. Here we can see how MOEA/DD have solutions both spread and at extremes points. This is an excellent aspect of performance and corroborates what happened for Hypervolume values. We can also see MOMBI2 performing well here with values less spread but near to optimal. That is the reason why MOMBI2 got statistically tied results with MOEA/DD in Hypervolume, but the less diverse solutions made it have worse results when compared to MOEA/DD regarding IGD+.

\subsection{Solution analysis}
\label{subsec:solutionanalysis}

In this section, we analyse the results from the problem perspective. Due to the fact MOEAs return a set of solutions, decision-makers have to choose which solution is more suitable to solve the problem. This decision is made by prioritising one or more objectives over the others. In our case, we chose to select from the $PF_{known}$ solutions which maximise a specific objective over the others. Table~\ref{tab:bestsolution} presents the chosen solutions for both two and five objective problems.

\begin{table*}[!htbp]
	\centering
	\caption{Best solutions found for each objective}
	\label{tab:bestsolution}
	\resizebox{1\textwidth}{!}{%
		\begin{tabular}{lcc}
			\toprule
			Best on & Two Objectives & Five Objectives                                                    \\
			\midrule
			Obj1 & {[}0,99999582, 4,43436123{]} & {[}0,99958084, 6,53043478, 0,20010116, 0,61349333, 0,98719498{]} \\
			Obj2 & {[}0,99992542, 7{]}          & {[}0,75522445, 7, 0,31699867, 0,43889046, 0,99860216{]}          \\
			Obj3 & -             & {[}-0,60228959, 6,91304348, 0,99902543, 0,52972268, 0,01606242{]} \\
			Obj4 & -             & {[}-0,48832059, 2,82608696, 0,85923814, 1,05, 1{]}                \\
			Obj5 & -             & {[}-0,22667412, 6,22608696, 0,64804175, 1,03502481, 1{]}      \\
			\bottomrule
		\end{tabular}
	}
\end{table*}

 In general, to analyze the results from the problem perspective we check the top 10 optimum results for each objective and see how prioritising it would affect the other objectives numbers. We report whether the other objectives numbers are low, too low or close to their best achieved numbers when they are prioritised. Also, if no specific pattern (high or low numbers) is detected, this may be an indicator that the objectives are independent. Moreover, for more precise comparison, if there is high discrepancy in the comparison the average of all values of the comparable objective is calculated to measure whether the achieved numbers are below or above average.

\subsubsection{The two objectives problem:}
    From the results of the two objectives problem optimisation, we can find that all the results of Obj1 (Equality) are satisfactory as the discrepancy between the best equality result (0.99999582) and the 10$^{th}$ result is just 0.007\%. However, the fairness (Obj2) values decline  drastically from the first solution to the last. This is because the number of evaders is 
    small compared to the total citizens and pushing all evaders to wealth group $g_1$ (Obj2) can have very slight impact on equality. In other words, optimum results of Obj1 can be achieved in case of prioritising Obj2. While optimising equality (Obj1) does not guarantee that all evaders end up in $g_1$ so Obj2 is not optimised in case of Obj1 prioritisation.
 \subsubsection{The five objectives problem:}
    \textbf{prioritising Equality (Obj1)}: This objective maximisation aims to reach having citizens with equal wealth, i.e. not favoring any wealth group. Results show that as equality is maximised the values of Obj3 (\emph{Wealth} of wealth group $g_5$) and Obj4 (\emph{Gained Amount} of wealth group $g_4$) are negatively affected. This is because Obj3 and Obj4 tend to maximise the wealth amount of a specific wealth group over the others. The impact on Obj3 is huge as its average result in case of the optimum values of Obj1 is approximately 0.2. However, Obj4 does not have values lower than 0.6 in case of Obj1 optimum value. This is because Obj4 does not directly affect the wealth of wealth group $g_4$ as it aims to maximise its portion from the gained amount. While the results of Obj2 and Obj5, in case of the optimum values of Obj1, do not show any specific pattern. Having equal wealth does not guarantee the number of evaders in wealth group $g_1$ (Obj2). Also, having equal wealth is not directly related to defining the collect rate of wealth group $g_1$ (Obj5).\\
    \textbf{prioritising Fairness (Obj2)}: On one hand, optimising Obj2 has huge negative impact on Obj3 similar to Obj1. This can be a consequent effect from having a higher catch rate $n3$ and fine $n4$ to be paid by evaders so they would be pushed to wealth group $g_1$. Accordingly, the wealth of evaders in wealth group $g_5$ will be decreased which is opposite to Obj3. Also, Obj4 is negatively affected in case of Obj2 optimum results, having an approximate result of 0.7 which is above its mean value. 
    On the other hand, Obj1 can have close to optimum performance, in case of the optimum values of Obj2. The difference between Obj1 highest value with Obj2 and the optimum value of Obj1 is only 0.0256\%. Although Obj2 tend to push most of the citizens to the middle wealth group to have equal wealth the number of evaders in this problem is not effective compared to the total number of citizens (only 0.05\%). So, if all evaders were pushed to wealth group $g_1$ (Obj2), Obj1 can still have close to optimum performance.
    While the results of Obj5, in case of the optimum values of Obj2, do not have specific pattern as the collect rate of $g_1$ is not directly related to the number of evaders in $g_1$. \\
    \textbf{prioritising maximum wealth of a specific wealth group (Obj3)}: Obj3 optimisation has negative effect on Obj1, Obj2 and Obj4 but with different levels. The effect on Obj1 is the worst, because supporting the wealth of specific group ($g_5$) is totally opposing the equality objective Obj1). 
    Also, maximising the wealth of the whole wealth group ($g_5$) including evaders (Obj3), does not support Obj2 which is to maximise the number of evaders in ($g_1$) and so decrease there wealth regardless their original wealth groups. 
    Obj4 results decrease but are still above the mean values. This is because increasing the new wealth of $g_5$ as a total percentage from the total wealth means that the new wealth of the rest of the wealth groups will be minimised. Consequently, the gained amount of $g_4$ (Obj4) will decrease.
     While the results of Obj5, in case of the optimum values of Obj3, do not have specific pattern as the collect rate of $g_1$ is not directly related to wealth of $g_5$. 
    \\
    \textbf{prioritising gained amount maximisation for one of the wealth groups (Obj4)}
    Obj4 optimisation has huge negative effect on Obj1 and Obj2, and has negative impact on Obj3. Equality (Obj1) is not supported by prioritising the gained amount (subsequently the wealth) of $g_4$ (Obj4).
     Also, maximising the gained amount of the whole wealth group ($g_4$) including evaders (Obj3), does not support Obj2 which is to maximise the number of evaders in ($g_1$) and so decrease there gained amount regardless their original wealth groups. Obj3 best value in case of the optimum values of Obj4 is reduced by 10.65\% compared to Obj3 optimum value. Increasing the gained amount of $g_4$ means that $g_4$ aims to have higher wealth which would affect the portion of wealth of $g_5$ (Obj3). 
     Contrary, Obj4 can still reach optimum results when prioritising Obj5, because the gained amount of $g_4$ (Obj4) maximisation do not contradict with minimising the collect rate of $g_4$(Obj5).
    \\
    \textbf{prioritising reducing the collected rate of a specific wealth group}\textbf{ (Obj5)}:
    Obj5 optimisation has negative effect on Obj1 and Obj3 but with different levels. Aiming to have a collect rate equal or near to zero (Obj5) would significantly decrease the equality (Obj1) because this means that this group is not treated equally like the rest of the groups. 
    The average of the results of Obj3, in case of prioritising Obj5, is still above its mean. A reasonable performance of Obj3 can be achieved as its best value in case of Obj5 prioritization is only reduced by 16.22\%.
    Obj5 can still reach close to optimum results when prioritising Obj4 as the difference between the best value of Obj4 in case of Obj5 optimum results and the optimum value of Obj5 is only 0.128\% .This is because the gained amount of $g_4$ (Obj4) maximisation do not contradict with minimising the collect rate of $g_4$(Obj5).
     While the results of Obj2, in case of the optimum values of Obj5, do not have specific pattern as the collect rate of $g_1$ (Obj5) is not directly related to the number of evaders in $g_1$ (Obj2).

As it can be seen from the analysis of the results from the problem perspective, prioritising any of the objectives can impact the other objectives, and have its pros and cons. Basically, the optimum solution can be further chosen based on the decision-maker priorities, and this results highlights the importance of having a good understanding of values dependencies. To do so, this model can be further extended by a reasoner that uses different reasoning techniques to recommend a final decision base on system wide priorities. 
\section{Discussion}\label{sec:Discussion}
After reviewing the results, it can be noted that NOA was successfully able to 
\begin{itemize}
    \item \textbf{Optimize multiple values regardless of their compatibility:} it is common in complex systems and large communities to have different values and each agent's values can be distinct from the other agents, so it is essential to be able to optimally compromise all of the values. NOA was able to reach this by turning the five values (equality, fairness, wealth, gained amount and collect portion) to objectives and formalising them as a multi-objective problem to be optimised. The produced pareto front set ensures having the optimum solutions for the combination of all objectives regardless of their compatibility. 
    \item \textbf{Select the best set of norms for heterogeneous groups of agents:} NOA was also able to select the best set of norms for heterogeneous groups of agents by using the parameterized norms techniques that allow each group of agents to have norms with different values. 
    \item \textbf{Align independent sets of values and norms:} by defining the norms set as the decision variables set in the MOP and the values set as the objectives to be optimised NOA was able to separate the concept of norms and values and having independent sets.
\end{itemize}

\section{Conclusion}\label{sec:conclusion}
In this paper, we proposed NAO, a multi-value promotion model that uses multi-objective evolutionary algorithms to produce the optimum parametric set of norms that is aligned with multiple simultaneous shared values and heterogeneous agents groups' values. It facilitates aligning multiple system values simultaneously with various norms. Moreover, it aligns values that may be incompatible such as fairness and equality (as ensuring fairness does not necessarily support equality), and aligns values of heterogeneous agents. More importantly, we show how different algorithms can have different performance in this domain by analysing the performance of four evolutionary algorithms (NSGA-II, MOEA/DD, SPEA2, and MOMBI2), and highlight the need for understanding the dependencies of various values in the system and how they might impact each other.
There are two directions for extending our work in the future. First, as performed in~\cite{deCarvalhoFatemeh2022}, adding an automatic reasoner to our model recommend a solution from the optimised solution set using different reasoning techniques. Second, to have an online mechanism that enables multi-value alignment replacing the current offline approach. For example, a hyper-heuristic approach could be employed on this purpose \cite{deCarvalho2019}. Subsequently, the limitation of only including a limited and static number of pre-defined norms needs to be addressed by on-line norm emergence techniques. Accordingly, more advanced scenarios of norms and values alignment can be further used for evaluation.

\bibliographystyle{plain}
\bibliography{sn-bibliography.bib}

\begin{thebibliography}{10}

\bibitem{pdfRefOlacir}
S.~F. Adra.
\newblock {\em Improving Convergence, Diversity and Pertinency in
  Multiobjective Optimisation}.
\newblock PhD thesis, Department of Automatic Control and Systems Engineering,
  The University of Sheffield, 2007.

\bibitem{aydougan2021nova}
Reyhan Aydo{\u{g}}an, {\"O}zg{\"u}r Kafali, Furkan Arslan, Catholijn~M Jonker,
  and Munindar~P Singh.
\newblock Nova: Value-based negotiation of norms.
\newblock {\em ACM Transactions on Intelligent Systems and Technology (TIST)},
  12(4):1--29, 2021.

\bibitem{bench2017norms}
Trevor Bench-Capon and Sanjay Modgil.
\newblock Norms and value based reasoning: justifying compliance and violation.
\newblock {\em Artificial Intelligence and Law}, 25(1):29--64, 2017.

\bibitem{surveyMetaHeuristics}
I.~Boussaid, J.~Lepagnot, and P.~Siarry.
\newblock A survey on optimization metaheuristics.
\newblock {\em Information Sciences}, 237:82 -- 117, 2013.

\bibitem{WFGPisa}
L.~Bradstreet, L.~Barone, L.~While, S.~Huband, and P.~Hingston.
\newblock Use of the {WFG} toolkit and {PISA} for comparison of {MOEAs}.
\newblock In {\em 2007 IEEE Symposium on Computational Intelligence in
  Multi-Criteria Decision-Making}, pages 382--389, Honolulu, USA, April 2007.
  IEEE.

\bibitem{coelho2021review}
Duarte Coelho, Ana Madureira, Ivo Pereira, and Ramiro Gon{\c{c}}alves.
\newblock A review on moea and metaheuristics for feature-selection.
\newblock In {\em International Conference on Innovations in Bio-Inspired
  Computing and Applications}, pages 216--225. Springer, 2021.

\bibitem{deCarvalhoFatemeh2022}
Vinicius~Renan de~Carvalho and Fatemeh Golpayegani.
\newblock Satisfying user preferences in optimised ridesharing services: A
  multi-agent multi-objective optimisation approach.
\newblock {\em Applied Intelligence}, 52(10):11257–11272, aug 2022.

\bibitem{deCarvalho2019}
Vinicius~Renan de~Carvalho, Kate Larson, Anarosa Alves~Franco Brand\~ao, and
  Jaime Sim~ao Sichman.
\newblock Applying social choice theory to solve engineering multi-objective
  optimization problems.
\newblock {\em Journal of Control, Automation and Electrical Systems (JCAE)},
  31(6):119--128, February 2020.

\bibitem{nsgaII}
K.~Deb, A.~Pratap, S.~Agarwal, and T.~Meyarivan.
\newblock A fast and elitist multiobjective genetic algorithm: {NSGA-II}.
\newblock {\em IEEE Trans. Evol. Comput.}, 6(2):182--197, Apr 2002.

\bibitem{ghanadbashi2022using}
Saeedeh Ghanadbashi and Fatemeh Golpayegani.
\newblock Using ontology to guide reinforcement learning agents in unseen
  situations.
\newblock {\em Applied Intelligence}, 52(2):1808--1824, 2022.

\bibitem{giugni1912variabilita}
Corrado Giugni.
\newblock {\em Variabilit{\`a} e mutabilit{\`a}: Contributo allo Studio delle
  Distribuzioni e delle Relazioni Statistiche}.
\newblock 1912.

\bibitem{golpayegani2016multi}
Fatemeh Golpayegani, Ivana Dusparic, Adam Taylor, and Siobh{\'a}n Clarke.
\newblock Multi-agent collaboration for conflict management in residential
  demand response.
\newblock {\em Computer Communications}, 96:63--72, 2016.

\bibitem{heidari2019agents}
Samaneh Heidari, Nanda Wijermans, and Frank Dignum.
\newblock Agents with dynamic social norms.
\newblock In {\em International Workshop on Multi-Agent Systems and Agent-Based
  Simulation}, pages 112--124. Springer, 2019.

\bibitem{mombiII}
Raquel Hern\'{a}ndez~G\'{o}mez and Carlos~A. Coello~Coello.
\newblock Improved metaheuristic based on the r2 indicator for many-objective
  optimization.
\newblock In {\em Proc. of the 2015 Annual Conf. on Genetic and Evolutionary
  Computation}, GECCO '15, pages 679--686, New York, NY, USA, 2015. ACM.

\bibitem{IGDPlus}
Hisao Ishibuchi, Hiroyuki Masuda, Yuki Tanigaki, and Yusuke Nojima.
\newblock Modified distance calculation in generational distance and inverted
  generational distance.
\newblock In Ant{\'o}nio Gaspar-Cunha, Carlos Henggeler~Antunes, and
  Carlos~Coello Coello, editors, {\em Evolutionary Multi-Criterion
  Optimization}, pages 110--125, Berlin Heidelberg, 2015. Springer
  International Publishing.

\bibitem{jaszkiewicz2018improved}
Andrzej Jaszkiewicz.
\newblock Improved quick hypervolume algorithm.
\newblock {\em Computers \& Operations Research}, 90:72--83, 2018.

\bibitem{moeaddd}
Ke~Li, Kalyanmoy Deb, Qingfu Zhang, and Sam Kwong.
\newblock {An Evolutionary Many-Objective Optimization Algorithm Based on
  Dominance and Decomposition}.
\newblock {\em IEEE Trans. Evol. Comput.}, 19(5):694--716, oct 2015.

\bibitem{luke2016population}
Sean Luke.
\newblock Population methods.
\newblock {\em Essentials of Metaheuristics (second ed.). Lulu}, 2016.

\bibitem{originalProblem}
Nieves Montes and Carles Sierra.
\newblock {\em Value-Guided Synthesis of Parametric Normative Systems}, page
  907–915.
\newblock International Foundation for Autonomous Agents and Multiagent
  Systems, Richland, SC, 2021.

\bibitem{nagymulti}
Muhammad Nagy, Yasser Mansour, and Sherif Abdelmohsen.
\newblock Multi-objective optimization methods as a decision making strategy.
\newblock {\em International Journal of Engineering Research and Technology},
  9(3), 2020.

\bibitem{nahian2020learning}
Md~Sultan~Al Nahian, Spencer Frazier, Mark Riedl, and Brent Harrison.
\newblock Learning norms from stories: A prior for value aligned agents.
\newblock In {\em Proceedings of the AAAI/ACM Conference on AI, Ethics, and
  Society}, pages 124--130, 2020.

\bibitem{jMetal}
A.~J. Nebro, J.~J. Durillo, and M.~Vergne.
\newblock Redesigning the jmetal multi-objective optimization framework.
\newblock In {\em Proc. of the Companion Publication of the 2015 Annual Conf.
  on Genetic and Evolutionary Computation}, GECCO Companion '15, pages
  1093--1100, New York, NY, USA, 2015. ACM.

\bibitem{noothigattu2019teaching}
Ritesh Noothigattu, Djallel Bouneffouf, Nicholas Mattei, Rachita Chandra,
  Piyush Madan, Kush~R Varshney, Murray Campbell, Moninder Singh, and Francesca
  Rossi.
\newblock Teaching ai agents ethical values using reinforcement learning and
  policy orchestration.
\newblock {\em IBM Journal of Research and Development}, 63(4/5):2--1, 2019.

\bibitem{ochoa2020multi}
Gabriela Ochoa, Lee~A Christie, Alexander~E Brownlee, and Andrew Hoyle.
\newblock Multi-objective evolutionary design of antibiotic treatments.
\newblock {\em Artificial intelligence in medicine}, 102:101759, 2020.

\bibitem{qu2018survey}
Bo-Yang Qu, YS~Zhu, YC~Jiao, MY~Wu, Ponnuthurai~N Suganthan, and Jing~J Liang.
\newblock A survey on multi-objective evolutionary algorithms for the solution
  of the environmental/economic dispatch problems.
\newblock {\em Swarm and Evolutionary Computation}, 38:1--11, 2018.

\bibitem{riad2022normative}
Maha Riad and Fatemeh Golpayegani.
\newblock A normative multi-objective based intersection collision avoidance
  system.
\newblock In {\em Agents and Multi-Agent Systems: Technologies and Applications
  2022}, pages 289--300. Springer, 2022.

\bibitem{riad2022run}
Maha Riad and Fatemeh Golpayegani.
\newblock Run-time norms synthesis in multi-objective multi-agent systems.
\newblock In {\em International Workshop on Coordination, Organizations,
  Institutions, Norms, and Ethics for Governance of Multi-Agent Systems}, pages
  78--93. Springer, 2022.

\bibitem{rodriguez2022instilling}
Manel Rodriguez-Soto, Marc Serramia, Maite Lopez-Sanchez, and Juan~Antonio
  Rodriguez-Aguilar.
\newblock Instilling moral value alignment by means of multi-objective
  reinforcement learning.
\newblock {\em Ethics and Information Technology}, 24(1):1--17, 2022.

\bibitem{serramia2021dominant}
Marc Serramia, Maite L{\'o}pez-S{\'a}nchez, Stefano Moretti, and Juan~Antonio
  Rodr{\'\i}guez-Aguilar.
\newblock On the dominant set selection problem and its application to value
  alignment.
\newblock {\em Autonomous Agents and Multi-Agent Systems}, 35(2):1--38, 2021.

\bibitem{serramia2020qualitative}
Marc Serramia, Maite Lopez-Sanchez, and Juan~A Rodriguez-Aguilar.
\newblock A qualitative approach to composing value-aligned norm systems.
\newblock In {\em Proceedings of the 19th International Conference on
  Autonomous Agents and MultiAgent Systems}, pages 1233--1241, 2020.

\bibitem{serramia2018exploiting}
Marc Serramia, Maite L{\'o}pez-S{\'a}nchez, Juan~A Rodr{\'\i}guez-Aguilar,
  Javier Morales, Michael Wooldridge, and Carlos Ansotegui.
\newblock Exploiting moral values to choose the right norms.
\newblock In {\em Proceedings of the 2018 AAAI/ACM Conference on AI, Ethics,
  and Society}, pages 264--270, 2018.

\bibitem{serramia2018moral}
Marc Serramia, Maite Lopez-Sanchez, Juan~A Rodriguez-Aguilar, Manel Rodriguez,
  Michael Wooldridge, Javier Morales, and Carlos Ansotegui.
\newblock Moral values in norm decision making.
\newblock In {\em Proceedings of the 17th International Conference on
  Autonomous Agents and MultiAgent Systems}, pages 1294--1302, 2018.

\bibitem{sierra2019value}
Carles Sierra, Nardine Osman, Pablo Noriega, Jordi Sabater-Mir, and Antoni
  Perell{\'o}.
\newblock Value alignment: a formal approach.
\newblock 2019.

\bibitem{vazquez2012mixture}
JA~V{\'a}zquez-Rodr{\'\i}guez and S~Petrovic.
\newblock A mixture experiments multi-objective hyper-heuristic.
\newblock {\em Journal of the Operational Research Society}, 64(11):1664--1675,
  2012.

\bibitem{spea2}
E.~Zitzler, M.~Laumanns, and L.~Thiele.
\newblock {SPEA2}: Improving the strength pareto evolutionary algorithm for
  multiobjective optimization.
\newblock In {\em Evolutionary Methods for Design Optimization and Control with
  Applications to Industrial Problems}, pages 95--100. CIMNE, 2001.

\end{thebibliography}



\end{document}